\documentclass{article}
\usepackage{arxiv}

\usepackage{amsmath,amssymb,amsfonts,amsthm}
\usepackage{algorithm}
\usepackage{algpseudocode}
\usepackage{enumitem}
\usepackage{mathrsfs}
\usepackage{url}
\usepackage{graphicx}
\usepackage{placeins}
\usepackage{xcolor}
\usepackage{hyperref}  

\usepackage{todonotes}
\usepackage{booktabs}

\theoremstyle{plain}

\theoremstyle{definition}

\theoremstyle{remark}

\title{Inexpensive Optical Projection Tomography on a Mobile Phone Platform}

\author{
  Gennifer T. Smith \\
  Department of Engineering \\
  University of San Francisco
  \And
  James M. Sikes \\
  Department of Biology \\
  University of San Francisco
  \And
  Nicholas Dwork \\
  Department of Biomedical Informatics \\
  University of Colorado, School of Medicine
}
\begin{document}
\maketitle

\begin{abstract}
This work presents an inexpensive optical projection tomography (OPT) system built on a mobile phone platform for three-dimensional optical microscopy. The system uses an iPhone camera together with a low-cost commercial microscope lens attachment, a stepper motor for sample rotation, LED illumination, and custom 3D-printed components, with a total component cost of approximately 50 US dollars excluding the phone. To support system evaluation, we also developed a low-cost method for fabricating a zebrafish phantom by embedding fixed larvae in UV-cured resin. Camera calibration was performed using a checkerboard target, and effective magnification was estimated with images of a 1951 Air Force resolution target. Projection images acquired during sample rotation were converted to attenuation images and corrected for field nonuniformity. Each slice was reconstructed with filtered backprojection and the resulting slices were stacked into a 3D volume. The completed system achieved a resolution of 3.91 $\mu m$ and produced volumetric reconstructions in which anatomical features of the zebrafish phantom, including the spine, were clearly visible. These results demonstrate that mobile-phone-based OPT can provide accessible, portable, and low-cost 3D microscopy, with potential utility for education, field work, and resource-limited settings.
\end{abstract}


\section{Introduction}

Microscopy has been used for centuries throughout medicine and biology, including pathology, embryology, cellular biology, virology, bacteriology, and other microbiology.  Once extremely expensive and requiring rare components, the optical microscope is now an inexpensive device commonly found in high school classrooms.  Inexpensive extensions (approximately \$30) can convert a smartphone into a microscope with up to a 200X magnification (e.g., the 200x Smartphone Microscope by Apexel).  Whereas the structures of interest are often three-dimensional, traditional optical microscopy limits imaging to two dimensions.  Either light travels through the object, in which case the observer sees a projection of the sample across space, or light is reflected off of the surface and the observer only sees the near-side of the object.  To compensate for this lack of three-dimensional imaging with optical microscopy, one often slices the object very thinly and then images each slice independently.  This is very laborious, requires significant talent, and still only presents one slice at a time to the observer.  Three-dimensional optical microscopy techniques exist \cite{merchant2023three}, such as confocal microscopy \cite{nwaneshiudu2012introduction} and optical coherence tomography \cite{huang1991optical}; however, the machines for this type of imaging are many tens of thousands of US dollars as most of them rely on focusing light at an array of small locations within the sample. A lesser known type of 3D microscopy is Optical Projection Tomography (OPT) \cite{sharpe2004optical, cheddad2011image, walls2005correction}, which manipulates the orientation of the sample relative to the illumination source.  With OPT, a set of visible-band light projection images are acquired as the sample is rotated within the optical path.  Tomography is used to reconstruct a 3D object from these projections.  Whiel not as expensive as confocal microscopy or OCT, OPT is still expensive.  E.g., the system of reduced cost published in \cite{vallejo2019optij}, described as a "cheap open-source software hardware and software OPT system", requires approximately 3000 US dollars in components and is designed for relatively large-scale microscopy subjects (as evidenced by the 8 mm deep sample imaged and 50 $\mu$m resolution).

In this manuscript, we present an OPT system based on a mobile phone platform that requires only 50 US dollars in components.  The system uses the camera of a mobile phone.  The rest of the system is comprised of an additional lens system, a motor to rotate the sample, three LEDs, and set of 3D printed components.  Altogether, other than the iPhone, the system costs $50$ US dollars.  The system is designed for thinner samples with smaller features (resolution of 3.91 $\mu$m).  

Phantoms are extremely useful for testing imaging devices and systems \cite{pogue2006review, lee2015fabrication, smith2015automated, smith2016multimodal}.  In addition to the inexpensive OPT system, we a method of creating a zebrafish phantom for imaging with this microscope using UV-cured resin.  Together, this work demonstrates the possibilities of an extremely inexpensive OPT system based on a mobile phone platform.

\section{Methods}

All procedures on live animals were approved by the Institutional Animal Care and Use Committee (IACUC) at the University of San Francisco.  In this seciton, we first describe the creation of the zebrafish phantom, we then discuss the manufacturing of the imaging system, and we finish with a description of the volumetric reconstruction method.

\subsection{Zebrafish Phantom}
In \cite{lin2018rigid}, Lin et al. present a method of creating a zebrafish phantom for micro-computed tomography imaging.  We modified their method to create an analogous phantom for OPT imaging.

The method of \cite{lin2018rigid} embedded millimeter-scale samples in heat-curable acrylic resin within a polymide tube. For visible light optical imaging, the presence of a tube outside the resin adds additional distortion to the images. Additionally, most polymide tubes are colored, which further alters the ability to optically image the sample. To better facilitate optical imaging and the removal of the tube from the resin without scratching the sample, samples were embedded in a UV-curable curable resin using a silicone tube. The quick (approximately one minute) cure time of the resin prevents bubbles from forming in the sample, and the silicone tube easily releases from the resin allowing it to be pulled off without leaving marks.

\subsubsection{Larvae Preparation}
Wildtype Zebrafish embryos were allowed to develop for approximately 24 hours, at which point the larvae were euthanized using a solution of MS-222 (Tricaine Methanesulfonate) dosed at 200 mg/mL buffered with sodium bicarbonate (NaHCO3) to a pH of 7. The euthanized samples were immersed in 10\% chilled neutral buffered formalin (NBF) and fixed overnight at room temperature. The fixed samples were rinsed 3 times in 1X phosphate buffed saline (PBS) for 10 minutes. Samples were submerged in 35\% ethyl alcohol (EtOH), then 50\% EtOH, then 70\% EtOH for 20 minutes each at room temperature with gentle agitation. Next, samples were then submerged in 90\% EtOH, then 95\% EtOH for 30 minutes at room temperature with gentle agitation, followed by two rounds of 100\% EtOH for 30 minutes with gentle agitation. 

\subsubsection{Embedding}
Unstained samples were submerged in a 1:1 volume/volume (v/v) mixture of 100\% EtOH and UV-curable clear resin overnight at room temperature with gentle agitation. Samples were placed in an opaque box to prevent curing of the resin. Samples were then submerged in 100\% UV-curable resin for 2 hours at room temperature with gentle agitation and away from light. After 2 hours, the resin was replaced with fresh resin and the samples were incubated for 1 hour at room temperature with gentle agitation away from light. 

\begin{figure}[!htbp]
  \centering
  \includegraphics[width=0.9\linewidth]{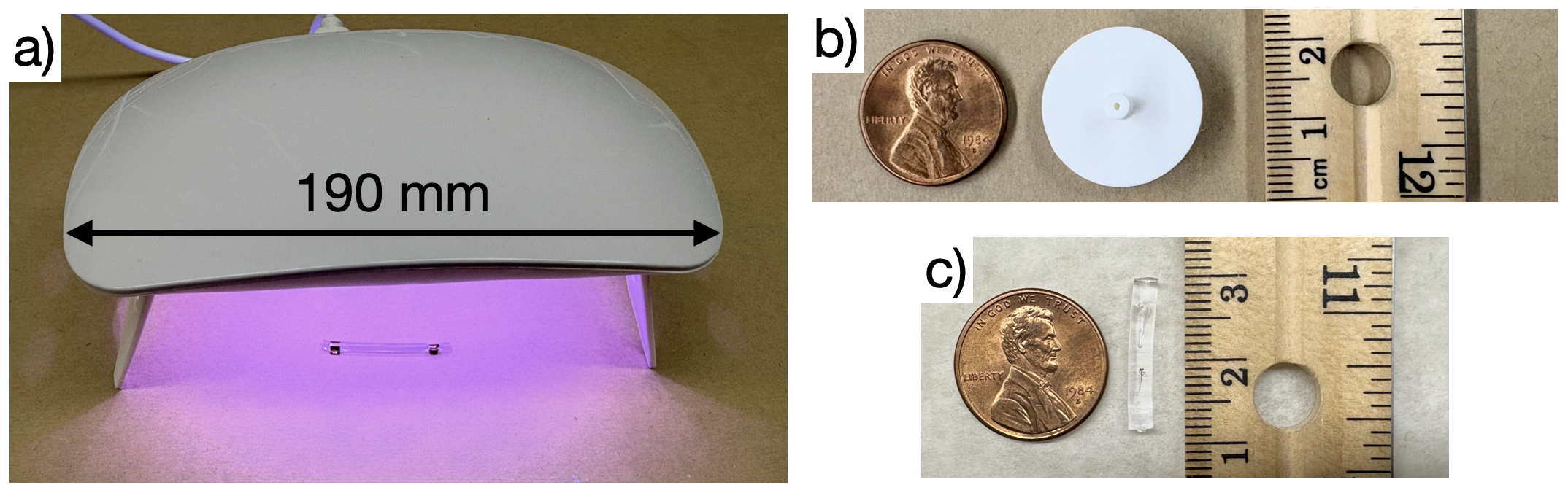}
  \caption{a) Example of crimp bead-sealed tube under UV light. Note that there isn't any zebrafish in this tube. b) 3D-printed holder used to embed smaller resin tube inside a larger resin tube. c) Completed zebrafish phantom next to a penny and a ruler for scale.}
  \label{fig:phantomFab}
\end{figure}

Samples in 100\% resin were transferred to a weight boat. The end of a $200$ $\mu$L micropipette tip was clipped until the 2.4 mm inner diameter (ID) silicone tube fit snugly in the end of the tip. The zebrafish was then pipetted slowly into the tube starting at the head. Care was taken to position the sample near the center of the tube to ensure the tubing above and below the specimen was filled with resin. Both ends of the silicone tube were sealed using crimp beads (often used in jewelry making) and the tube was placed horizontally under an ultraviolet (UV) light for $2$ minutes (Fig.~\ref{fig:phantomFab}a). 

After fully curing, the silicone tube was removed from the resin.
To embed the smaller resin sample inside a larger tube, a custom holder was 3D-printed (Figure \ref{fig:phantomFab}b). The holder was designed to hold the smaller sample in the center of the post while the larger silicone tube (3.0 mm ID) fit snugly around the outside of the post. 
The high viscosity of the UV-curable resin prevents filling the outer tube without bubbles forming. To reduce the viscosity without compromising the curability, a 1:4 v/v mixture of 100\% EtOH and UV-curable clear resin was used instead. This mixture was inserted into the larger tube with a pipette. Samples were cured for 2 minutes under UV light and the outer silicone tubing was removed. 

Due to mismatch in refractive indexes of the resin and surrounding air, and the resulting high internal reflection at the edges tube as viewed by the camera, the edges appear opaque when imaged (Fig.~\ref{fig:MotorAndLensAttachments}c). To ensure the sample is roughly centered in the resin so that it is not obscured by the opaque edges, zebrafish larvae were first embedded in a smaller-diameter tube (2.4 mm ID), allowed to fully cure, then embedded in a larger-diameter tube (3.0 mm ID) to ensure sufficient space between the zebrafish and outer edge of the resin. The diameter of the larger tube was chosen to balance sufficient distance between the opaque-appearing edges and the sample while also minimizing the total thickness of resin. The more resin that needed to be imaged through, the more distorted the images. 

The final zebrafish phantom is shown in Fig.~\ref{fig:phantomFab}c.

\subsection{Hardware Setup}
\begin{figure}[!htbp]
  \centering
  \includegraphics[width=0.6\linewidth]{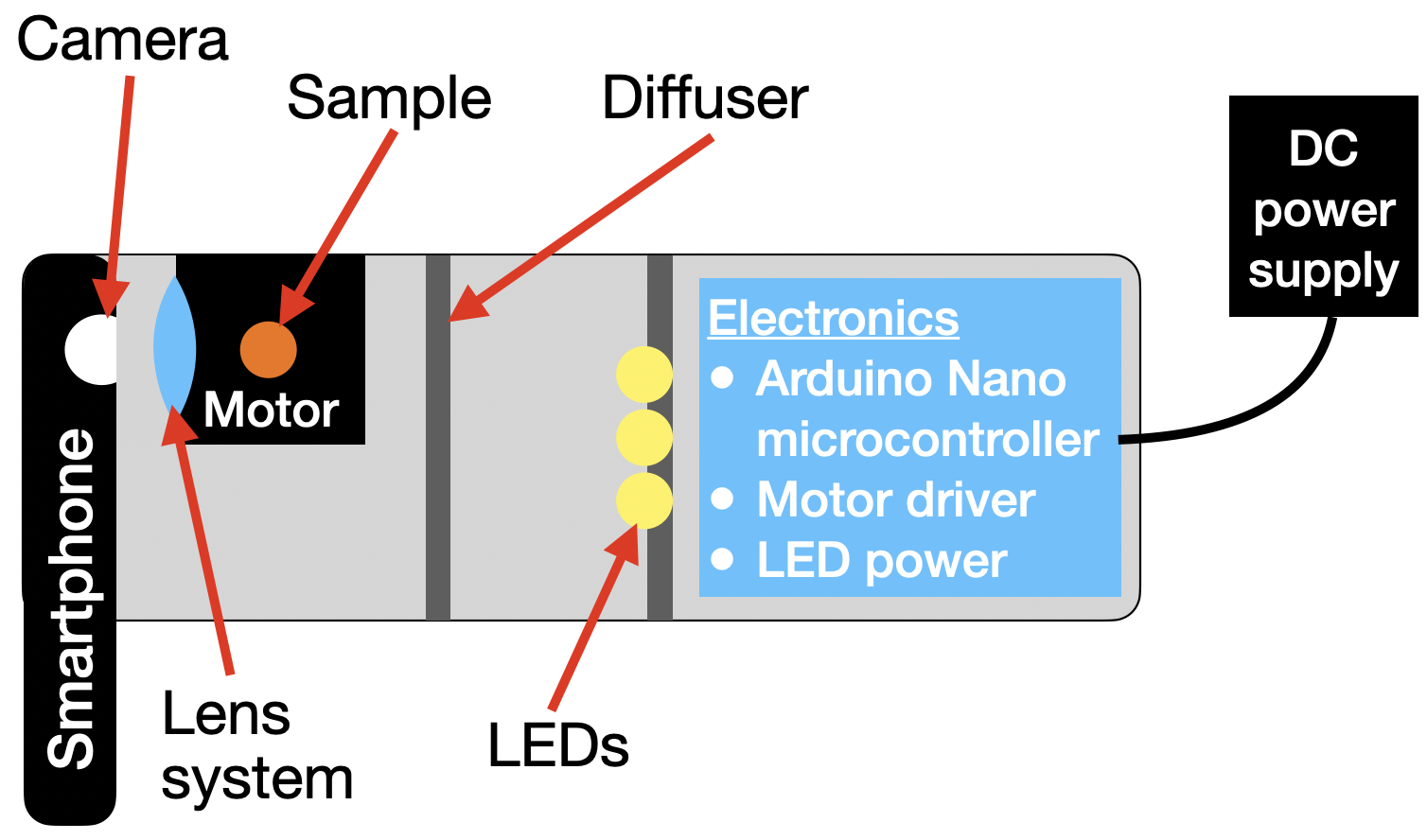}
  \caption{A schematic of the arrangement of components of the tomographic microscope. The electronics includes an Ardunio Nano, TMC2209 driver, and connections to power the LEDs. The light gray region shows the portion that is encased in a housing to hold all components in the proper location and to block external light.}
  \label{fig:SystemSchematic}
\end{figure}

In this section, we describe the creation and setup of the imaging system.  A schematic of the system is presented in Fig.~\ref{fig:SystemSchematic}.

\subsubsection{Motor and Electronics}
We used a NEMA 17 Stepper Motor (0.9 degree step size, 1.5 A) combined with a TMC2209 driver to rotate the sample. The combination of motor and driver were chosen for the low heat generation, the microstepping options, and the essentially non-existent jitter even at low speeds. The system was controlled using an Arduino Nano microntroller. To achieve a rotation rate of 1 degree per second, the Arduino was programmed to use 1/8 microstepping along with an appropriate delay between steps. The Arduino Nano and motor were both powered by the same a DC power supply. If there is easy access to a wall outlet, a 12 volt, 2 amp power adapter can be used. Alternatively, three 3.7 volt Li-ion rechargeable batteries can be connected in series and used as the power source. Both methods were tested, but the results presented in this manuscript were all generated with the 12 volt wall power adapter. To protect the Arduino Nano from overheating, the input voltage was stepped down from 12 volts to 5 volts using a voltage regulator.

\subsubsection{Imaging System}
All images were captured with an iPhone 16 using the Ultra Wide 12 megapixel camera. The ProShot 3rd party iPhone application manually controlled camera settings and collected images in Digital Native (DNG) format. To create a low-cost system, we took advantage of mass-produced cellphone microscope attachments. We utilized a combination consisting of a biconvex lens, an achromatic doublet, and a circular polarizing (CPL) filter (obtained as part of an APEXEL 200X Pocket Microscope Lens with CPL Filter). Three 8mm-diameter white LEDs (3 V, 250 mA) created sufficient illumination with low power consumption. To evenly distribute the illumination across the entire sample, a piece of white translucent acrylic (5 mm thickness) was placed between the sample and the LEDs (as seen in Fig. \ref{fig:SystemSchematic}). The acrylic was lasercut to the appropriate size. To block external light, all components were housed in a 3D-printed box made out of white polylactic acid (PLA) filament. All 3D-printed parts were made using a Bambu Lab X1-Carbon. 

\subsection{Volumetric Reconstruction}

The reconstruction pipeline converted a sequence of transmission images acquired at different rotation angles into a three-dimensional map of attenuation. This was done in four stages: estimation of the imaging geometry, correction and logarithmic transformation of the projection images, reconstruction of a two-dimensional cross-section from each horizontal detector line, and concatenation of those cross-sections into a volume. More specifically, for a fixed detector line, the intensity values recorded over all projection angles were treated as a sinogram and reconstructed to form one slice of the object. Repeating this procedure for all detector lines produced a stack of slices, which together formed the final three-dimensional reconstruction.  In the following subsections, we describe each stage in detail.

\subsection{Imaging Geometry}

A precise camera calibration is required for an accurate 3D reconstruction.  For this work, we modeled the optical system as a central projective camera \cite[Section 1.2]{hartley2003multiple}. The intrinsic camera parameters, which describe how directions in object space map to positions on the detector, must be known. These parameters are contained within the intrinsic camera matrix
\[
K =
\begin{bmatrix}
f_x & 0 & c_x \\
0   & f_y & c_y \\
0   & 0   & 1
\end{bmatrix},
\]
where $f_x$ and $f_y$ are the horizontal and vertical focal lengths in pixel units, respectively, and $(c_x,c_y)$ is the principal point.  (Note that a more general intrinsic camera matrix includes a non-zero $K_{1,2}$ value, called skew.  We assumed a skew of $0$ for this work \cite{medioni2004emerging}.) Once $K$ is known, detector coordinates can be related to ray direction, which is the information needed to interpret each projection as a fan-beam geometry.

\begin{figure}[!htbp]
  \centering
  \includegraphics[width=0.8\linewidth]{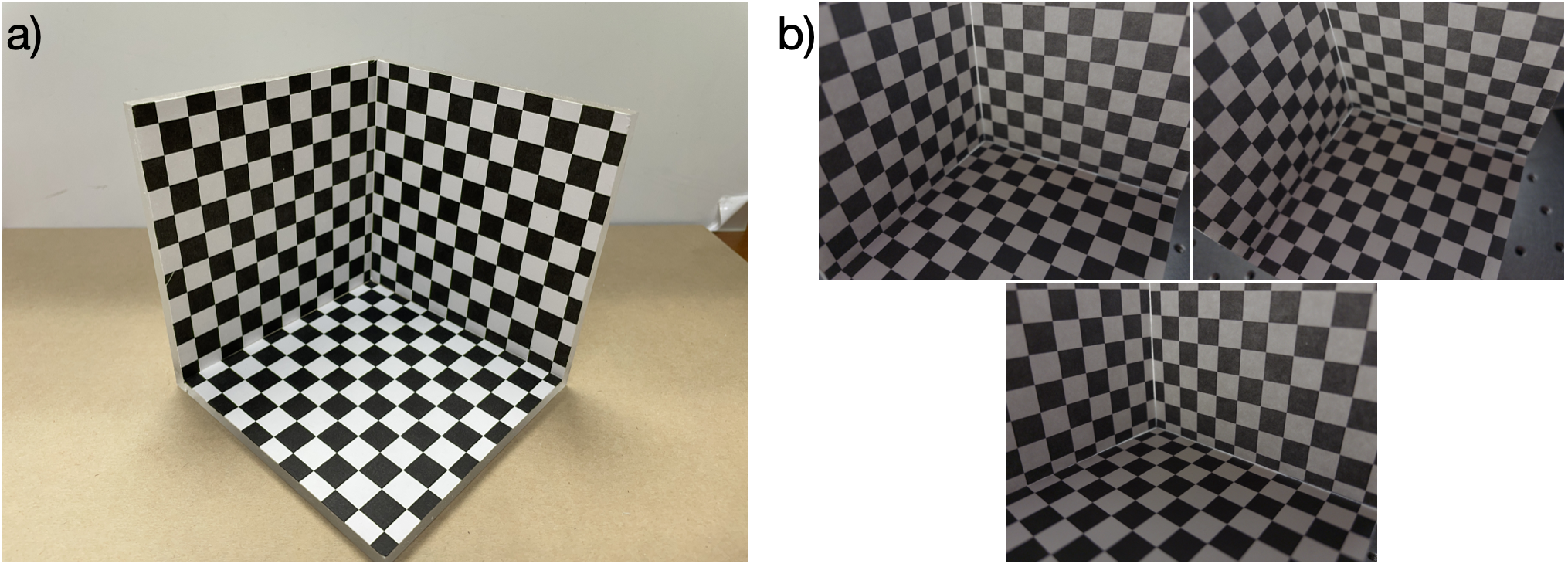}
  \caption{Images of the checkerboard pattern used to identify the intrinsic camera parameters of the camera: a) an image of the checkerboard target from afar; b) closer images used during the calibration process.}
  \label{fig:checkerboard}
\end{figure}

The intrinsic parameters were estimated from images of a planar checkerboard.  The checkerboard was created by plastic welding three pieces of laser-cut acrylic together to form a corner. A checkerboard pattern was printed on paper and calipers were used to ensure that the pattern consisted of true squares (in case of non-square pixels of the printer). The paper was attached to the acrylic using 3M 77 spray adhesive.

To calibrate the camera, $3$ or more square features are identified across the calibration images.  (Theoretically, $3$ square features in three different planes are enough to calibrate the camera.  However, due to noise, a larger number of square features is required for accurate calibration.  For the results of this manuscript, we identified 13 different squares.)
For each observed square, a homography $H$ is determined that maps the corners of the square in the image to the canonical square $(0,0),\ (1,0),\ (0,1),\ (1,1)$. Writing $H = [\,h_1\;\;h_2\;\;h_3\,]$, each homography yields the relations $h_1^{T}\omega h_2 = 0$ and $h_1^{T}\omega h_1 = h_2^{T}\omega h_2$, where $\omega$ is the image of the absolute conic \cite[Section 8.5]{hartley2003multiple}.
Under the zero-skew assumption used, these equations are linear in the unknown entries of $\omega$. The constraints from all measured squares were assembled into a single linear system and solved in a least-squares sense. The intrinsic matrix $K$ was then recovered from $\omega^{-1} = K K^{T}$ using a Cholesky factorization, and normalized so that \(K_{33}=1\) \cite{hartley2003multiple} \cite[Section 8.5]{hartley2003multiple}.

Before estimating the intrinsic matrix, the image coordinates of all corners were normalized to improve numerical conditioning of the homography and calibration computations. Let a point in detector pixel coordinates be written in homogeneous form as $\mathbf{x} = \begin{bmatrix} x & y & 1 \end{bmatrix}^T$.
The normalized coordinates were defined by $\tilde{\mathbf{x}} = T \mathbf{x}$, where $T$ is a similarity transformation that first translates the mean of the image points to the origin and then applies an isotropic scaling so that the mean Euclidean distance of all points to the origin is $\sqrt{2}$. Thus,
\[
T
=
\begin{bmatrix}
\alpha & 0 & -\alpha t_x \\
0 & \alpha & -\alpha t_y \\
0 & 0 & 1
\end{bmatrix},
\qquad
\alpha = \frac{\sqrt{2}}{\overline{d}},
\]
where $\alpha = \sqrt{2} / \overline{d}$ and $\overline{d}$ is the mean distance of the translated image points from the chosen center. Calibration was then carried out in these normalized image coordinates, producing an intrinsic matrix $\tilde{K}$.

If $\tilde{K}$ denotes the intrinsic matrix estimated in the normalized coordinate system, then the intrinsic matrix in detector pixel coordinates is obtained by: $K = T^{-1}\tilde{K}$.
After this back-transformation, $K$ was normalized by dividing each element of $K$ by $K_{33}$ (so that $K_{33}=1$).

\begin{figure}[!htbp]
  \centering
  \includegraphics[width=0.5\linewidth]{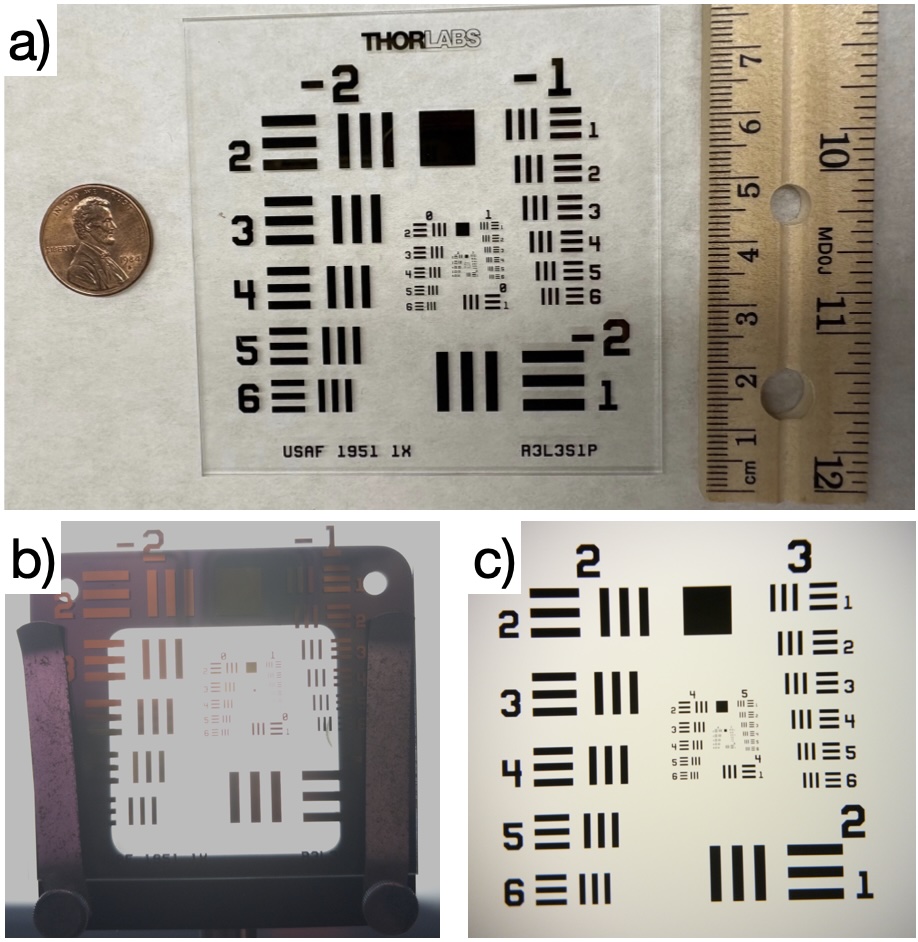}
  \caption{Images of the 1951 Air Force Target captured with an iPhone camera a) without and b) with the additional magnification lens.}
  \label{fig:airForceTarget}
\end{figure}

The checkerboard calibration was performed without the additional magnification lens used during OPT acquisition. The resulting calibration therefore provided the base camera geometry, but not the exact effective intrinsics of the final optical train. To account for this, the focal lengths from the bare-camera calibration was scaled by a magnification factor $M$.  Images of the 1951 Air Force target were used to identify this magnification factor (Fig.~\ref{fig:airForceTarget}a).  Specifically, images were taken of the Air Force target with and without the additional lens, as shown in Fig.~\ref{fig:airForceTarget}b,c. We also measured the distance from the camera center to the target for each image taken which were changed to maintain focus. The distances were significantly different since the distance to the plane of focus changed with the additional lens and the depth of focus was small.  With known lengths and widths for each line of the target, we were able to determine the horizontal angular extent of the each pixel.  The ratio of those angular extents yielded the magnification $M$.

If $K_{11}$ denotes the focal length in horizontal pixel units from the calibration and $f_{\mathrm{bare}}$ is the physical focal length of the bare camera (without the additional magnifying lens system), then the detector pixel pitch was estimated as $p = f_{\mathrm{bare}} / K_{11}$, and the effective focal length used for reconstruction was taken to be $f = M K_{11}$.

We made the approximation that the principal point remained unchanged with the addition of the magnification lens.

\subsection{Projection formation and field-intensity correction}

\begin{figure}[!htbp]
  \centering
  \includegraphics[width=0.8\linewidth]{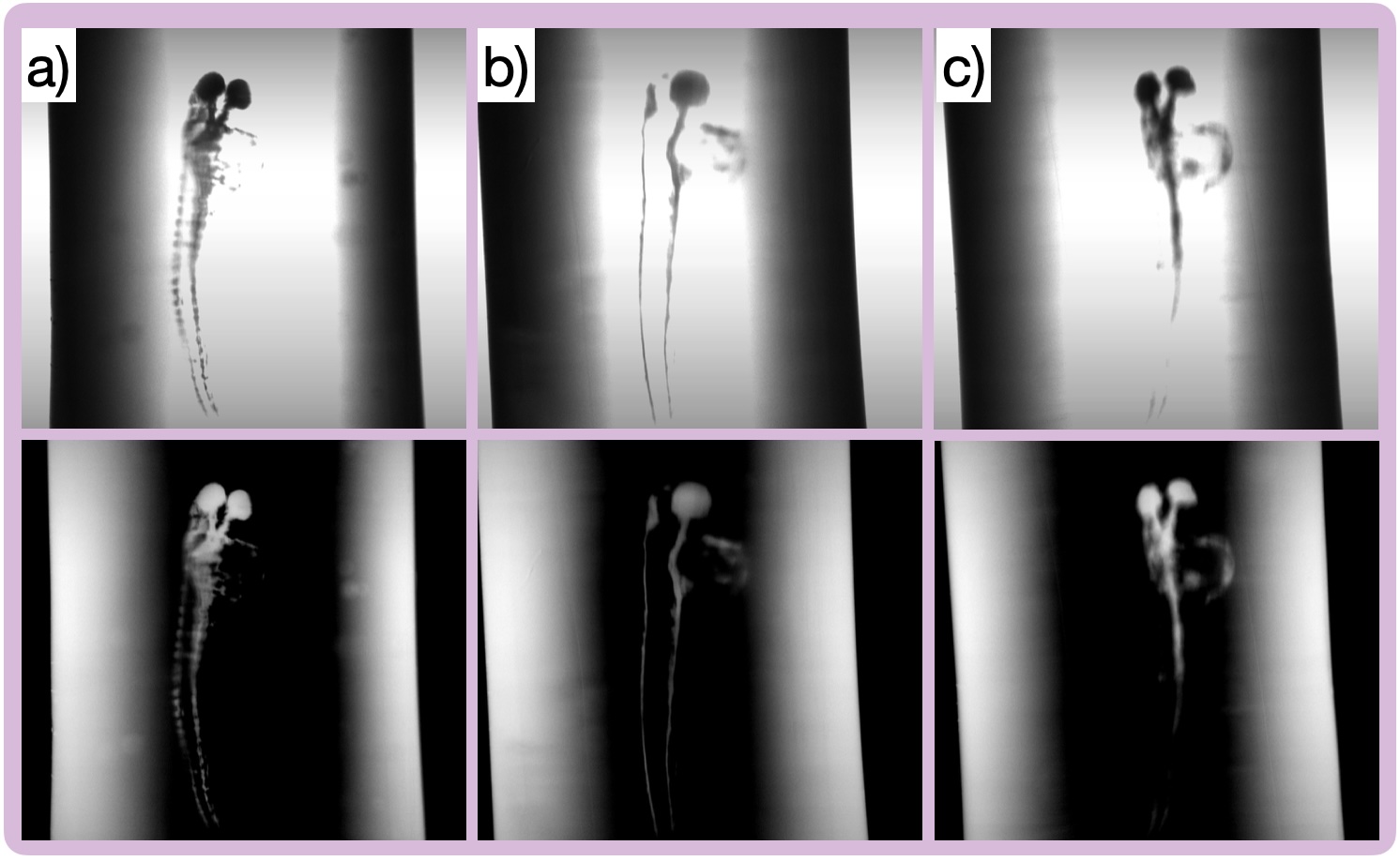}
  \caption{Top row: examples of the luminance projection images captured by the iPhone. Bottom row: corresponding attenuation projections.  Subimages a, b, and c show rotations of $0^\circ$, $45^\circ$, and $90^\circ$, respectively.}
  \label{fig:attnImages}
\end{figure}

Prior to 3D volume reconstruction, the raw projection images were converted into attenuation projections (Fig.~\ref{fig:attnImages}). This step had three purposes: to restrict the data to the detector region used for reconstruction, to correct for spatial variations in the illumination field, and to express the measurements as line integrals suitable for tomographic inversion.

Images were captured using the ProShot iPhone application in raw Digital Native (DNG) format.  Data was read in using Matlab's \verb|rawread| function and Bayer interpolation was accomplished using Matlab's \verb|demosaic| function.  The green channel was then isolated for further processing.  (The blue and red channels were discarded.)
Let \(I(u,v,\theta)\) denote the measured intensity at detector column \(u\), detector row \(v\), and projection angle \(\theta\).

The field-intensity correction was performed independently for each projection and was designed to remove slow spatial variation in the illumination across the detector. Rather than assuming a single scalar reference intensity, the incident field was estimated separately for each detector row using air pixels near the lateral borders of the detector. For each row \(v\), 40 columns from the left side and 40 columns from the right side of the valid detector support were pooled, and the reference intensity was taken as the row-wise median, $I_0(v,\theta) = \operatorname{median}_{u \in \Omega_{\mathrm{air}}} I(u,v,\theta)$, where \(\Omega_{\mathrm{air}}\) denotes border columns that lay outside the specimen. The resulting profile was smoothed with a moving-median filter. The estimate for each row was then replicated across detector columns to form a two-dimensional incident-field image $I_0(u,v,\theta) = I_0(v,\theta)$.

With this incident field, each projection was converted to a line-integral image with the Beer--Lambert transform,
\[
\mu(u,v,\theta) = -\log\!\left(\frac{I(u,v,\theta)}{I_0(u,v,\theta)}\right).
\]

After preprocessing, the data consisted of a three-dimensional stack of line-integral projections with dimensions $(\text{detector row}) \times (\text{detector column}) \times (\text{projection angle})$.
The key simplification used in the reconstruction was to treat each detector row as an independent two-dimensional fan-beam acquisition. In this view, a fixed detector row provides a sinogram over detector position and rotation angle, which can be rebinned into a standard parallel-beam representation before reconstruction.  This is an approximation that neglects the angle of the fame beam away from horizontal for rows that do not include the principal point.

Each fan-beam sample acquired at detector position \(u\) and view angle \(\theta\) was assigned to the parallel-beam coordinates $(s_u,\ \theta+\gamma_u)$.
A rebinned parallel-beam sinogram was then formed on a uniform $(s,\theta)$ grid using linear interpolation in this coordinate system. Once a detector row had been rebinned into a parallel-beam sinogram, it was reconstructed with filtered backprojection \cite{kak2001principles}.

The full three-dimensional volume was obtained by repeating the row-wise reconstruction for every detector row in the cropped projection stack. Each detector row therefore produced one reconstructed two-dimensional slice, and these slices were stacked in detector-row order to form the final OPT volume.

\section{Results}
\subsection{Hardware System}

\begin{figure}[!htbp]
  \centering
  \includegraphics[width=0.8\linewidth]{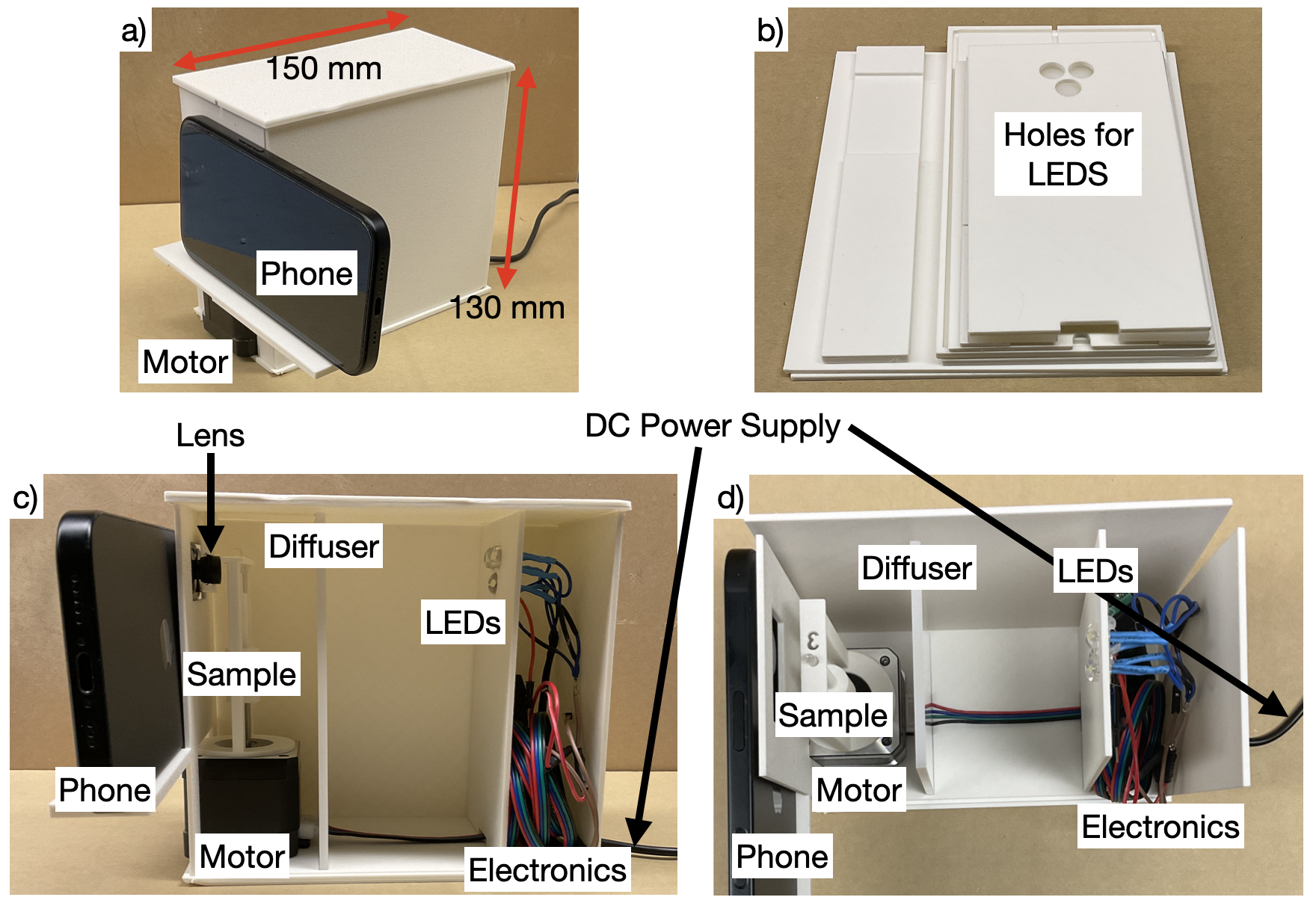}
  \caption{a) Outside view of 3D-printed box. b) Disassembled 3D-printed box to show how it can lay flat for shipping. c) Side view of the interior of the 3D-printed box and the various components of the OPT system; d) top-down view of the system. Note that the lens is not shown in the top-down view because the lens is held in place by attaching it to the top of the box. }
  \label{fig:FullBox}
\end{figure}

\begin{figure}[!htbp]
  \centering
  \includegraphics[width=0.8\linewidth]{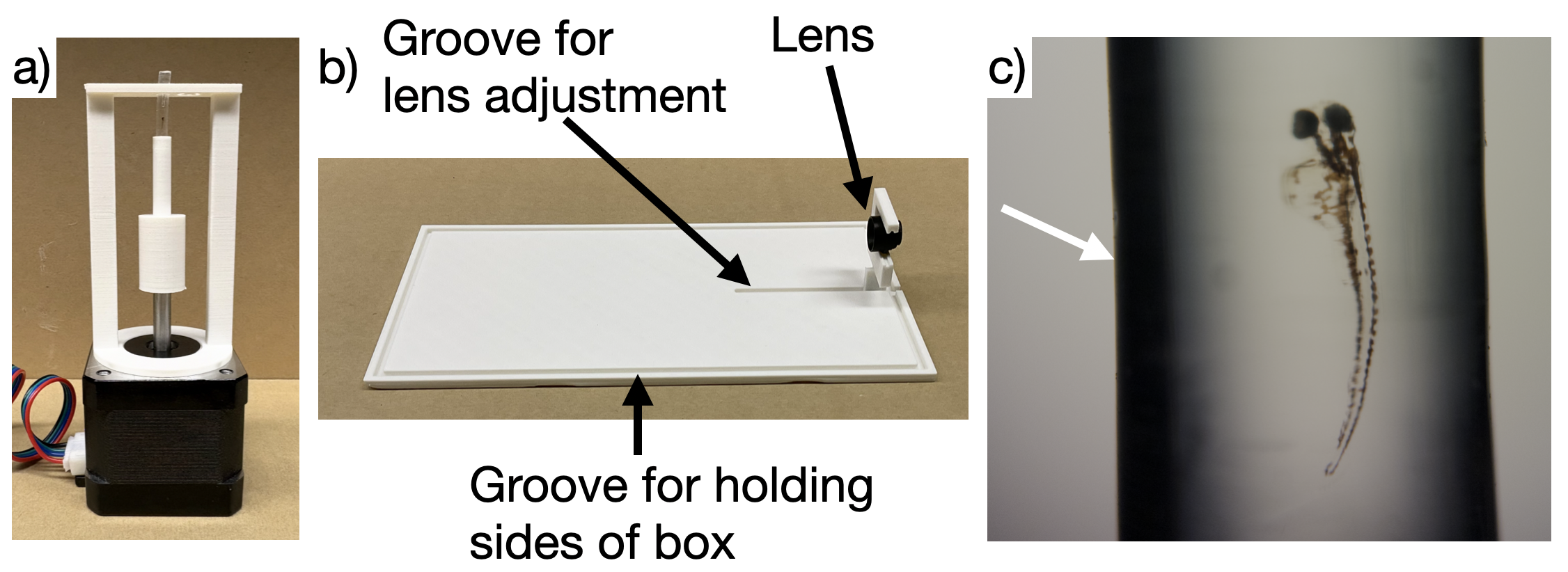}
  \caption{a) Motor attachment used to attach the sample tube to the NEMA motor shaft and keep the sample upright. The motor plus the attachment are 110 mm tall. b) The lid of the 3D-printed box and the attachment used to hold and position the lens the correct distance from the phone. The long side of the lid is 150 mm long. c) An image of the phantom as taken by the mobile phone microscope with white balance applied.  The white arrow indicates a region near the edge of the tube where the high internal reflection makes the tube appear opaque.}
  \label{fig:MotorAndLensAttachments}
\end{figure}

Figure \ref{fig:FullBox} shows complete system, including the 3D-printed box used to block external light and hold components in place.
To allow for sufficient working space if adjustments to the positions of the movable parts (motor and lens system) are needed, each face of the box and each interior surface were individually printed or laser cut. We added grooves to the bottom and top surfaces where the other parts could be inserted for a snug fit (Figure \ref{fig:MotorAndLensAttachments}b) shows an example of the groove in the lid (top surface). Note that the gap between some of the components in Figure \ref{fig:FullBox}b) are remedied when the lid is placed on top to hold the pieces together. The total cost of the components, including the 3D-printed and lasercut parts but excluding the cost of the phone, is 50 US dollars (Table ~\ref{table:partsList}), and the total weight is 652 grams. The Standard Tesselation (STL) Files for the 3D printed components can be found here: \verb|https://drive.google.com/drive/folders/1k6QyZamtAcp8LZGrEULDxjuBjFYj9u9f|.

\begin{table}[h!]
\centering
\caption{Components in the system and their associated costs, showing a total cost of 50 US dollars}
\label{table:partsList}
\begin{tabular}{l r}
\toprule
\textbf{Part} & \textbf{Cost (US Dollars)} \\
\midrule
APEXEL 200X Pocket Microscope Lens with CPL Filter & \$20.00 \\
NEMA 17 Stepper motor 0.9 degree & \$10.00 \\
TMC2209 V1.3 UART Stepper Motor Driver & \$5.00 \\
Nano Board ATmega328P & \$3.00 \\
8mm-diameter white LEDs (3 V, 250 mA) & \$0.50 \\
12 V adapter & \$7.50 \\
Acrylic & \$1.00 \\
3D printed material & \$3.00 \\
\midrule
\textbf{Total} & \textbf{\$50.00} \\
\bottomrule
\end{tabular}
\end{table}

\begin{figure}[!htbp]
  \centering
  \includegraphics[width=0.5\linewidth]{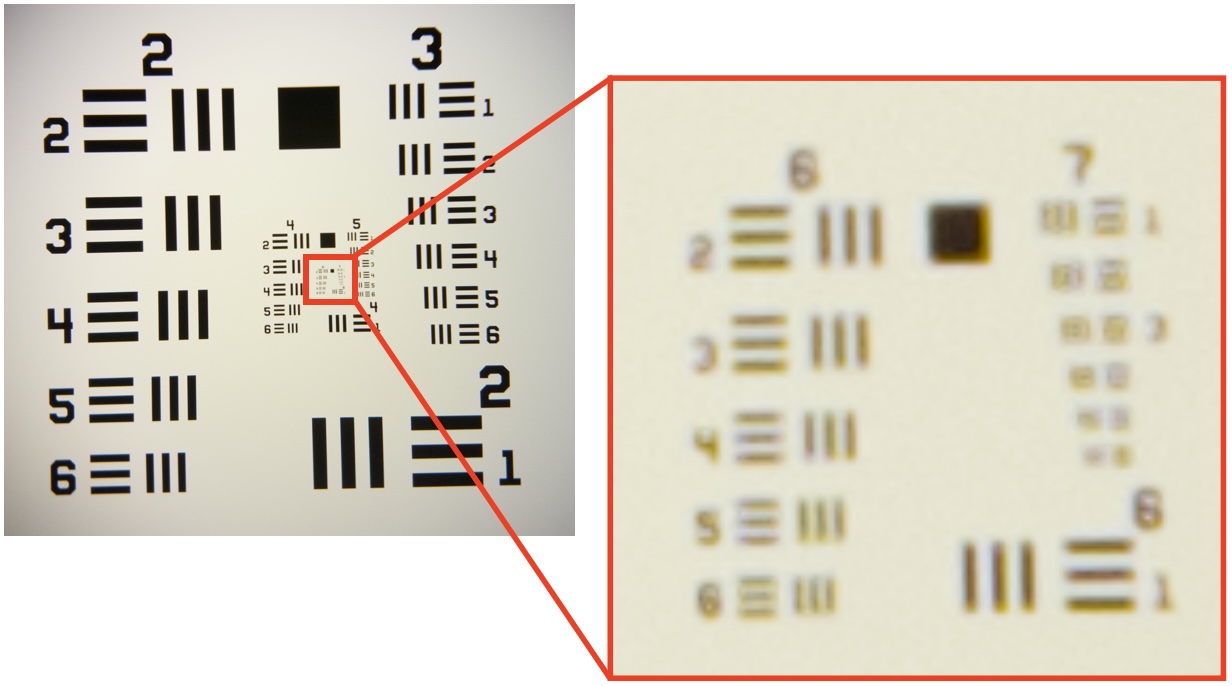}
  \caption{Image of the 1951 Air Force target captured by the smartphone microscope with a zoomed in region showing a resolution of 3.91 $\mu$m}
  \label{fig:systemRes}
\end{figure}

To attach the zebrafish phantom to the motor and ensure it remained vertical throughout data collection, we made a custom 3D-printed attachement as shown in Figure \ref{fig:MotorAndLensAttachments}a). Due to the mismatch in the size of the shaft of the NEMA 17 and the small distance required between the microscope lens system and the sample, the base of the attachment had to be wider than the top. Due the short focal length of the microscope lens system, there is a small margin of error for the distances between the smartphone camera, the microscope lens system, and the sample. To account for imperfections in the 3D printing process, we made a groove in the top of the 3D-printed box in which the removable lens holder could be attached (as seen in Figure \ref{fig:MotorAndLensAttachments}b)). The groove allows for the distance between the smartphone camera and lens to be adjusted. The front of the box also has an opening for the motor, which can be slid back and forth to adjust the distance between the sample the microscope lens. Figure \ref{fig:MotorAndLensAttachments}c) shows an image of the phantom taken using the motor attachment and microscope lens described. The full system, including the illumination with LEDs and a diffuser, has a resolution of 3.91 $\mu$m (as shown in Figure \ref{fig:systemRes}

\subsection{Reconstructions}
All images were captured with an iPhone 16 using the Ultra Wide 12 megapixel camera.
3rd party app (ProShot) to manually control camera settings and collect Digital Native (DNG) format. All images were captured with the following parameters: focus of 0, ISO of 140, shutter speed 1/60 s.

\begin{figure}[!htbp]
  \centering
  \includegraphics[width=0.8\linewidth]{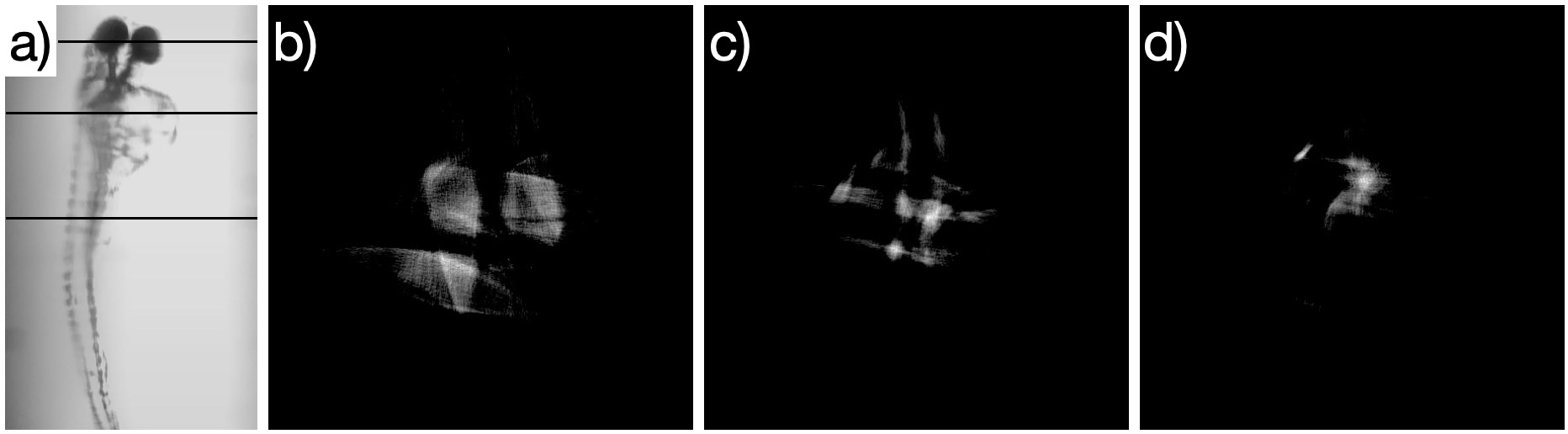}
  \caption{Reconstructions of individual slices of the zebrafish phantom.  a) A projection image of the zebrafish phantom with horizontal lines indicating the locations of the slices shown in subimages b, c, and d.  Subimages b, c, and d show reconstructions of the top, middle, and bottom lines, respectively.}
  \label{fig:slices}
\end{figure}

\begin{figure}[!htbp]
  \centering
  \includegraphics[width=0.8\linewidth]{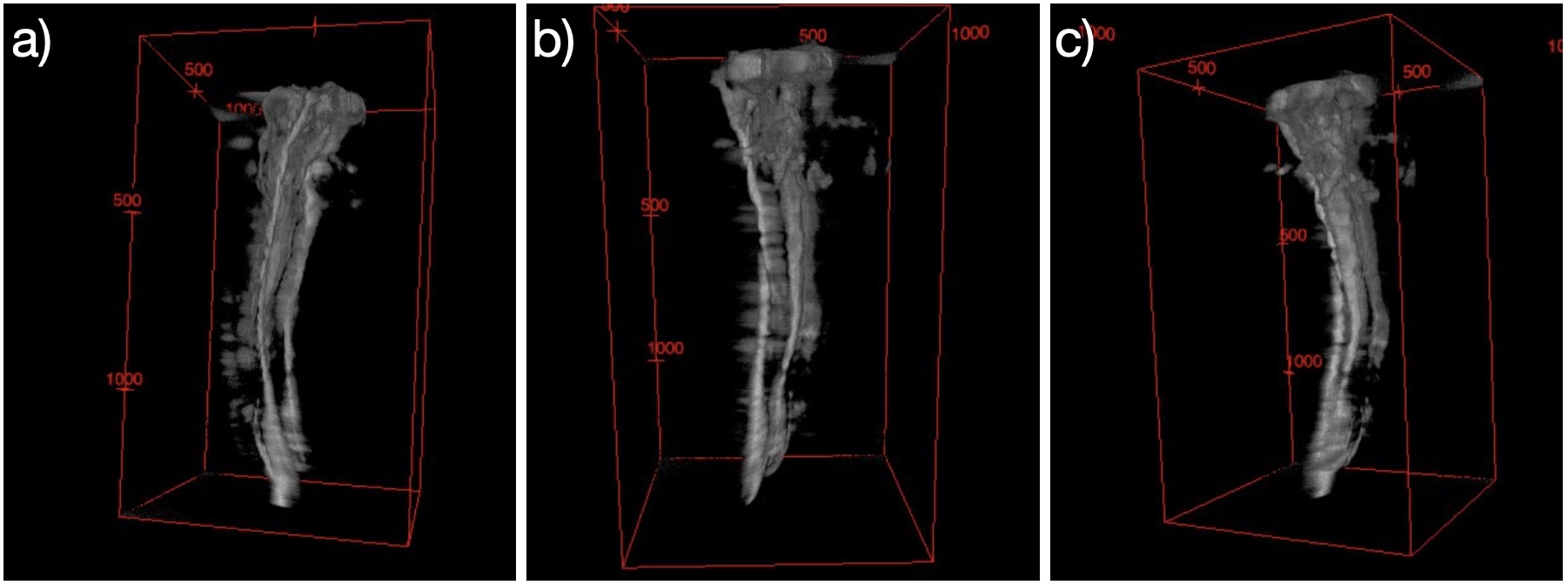}
  \caption{Visualizations of the reconstructed volume using 3D Viewer within the the FIJI ImageJ application.}
  \label{fig:fijiVis}
\end{figure}

Figure \ref{fig:slices} shows individual slices of the reconstruction of the zebrafish phantom.  Figure \ref{fig:fijiVis} shows a 3D rendering of the three-dimensional volume using the 3D Viewer plugin from within the Fiji ImageJ application.  For this rendering, voxels with low intensity were nearly transparent; the opacity of each voxel increased with intensity.  The spine of the zebrafish is clearly visible in the 3D volumetric reconstruction.

\section{Discussion and Conclusion}

This work demonstrates the feasibility of an inexpensive optical projection tomography system on a mobile phone platform. Indeed, the spine of the zebrafish is clearly rendered in Fig. \ref{fig:fijiVis}. Though reconstruction for the results presented in this manuscript was performed offline, the reconstruction algorithm is computationally simple and feasible for a mobile phone's computer.

The design of the system, including individually manufactured parts that are portable and easily assembled into a box on site and the option to power the entire system using batteries, makes it suitable for studies in resource-limited and remote locations. While the zebrafish phantom fabrication presented here required many steps over several days, it was done with the intention of creating a sample that would remain stable over a long period of time (months). For samples that only need to be viable over a short period of time (hours rather than months), a less laborious fabrication technique could be used. For example, the sample could be embedded in 100\% resin directly after fixation as opposed to being dried and infiltrated through a series of EtOH washes and overnight incubation steps. Therefore, the effort requird for sample preparation in a resource-limited or remote location could be significantly reduced.

Several approximations made during the reconstruction process may be limiting the accuracy of the reconstruction.  Specifically, we made the following approximations.  1) Light rays travel in straight lines.  This is not true when encountering a medium with a different index of refraction such as the tube of resin.  2) The principal point of the camera does not change when the lens system is included in the optical path. 3) We assumed that the axis of rotation was parallel to the vertical axis of the captured images. There may be an angle between these directions, which could be taken into account.  4) Each slice of data can be treated independently as a sinogram of projections.  In truth, the projections are angled for all but the slice that intersects the principal point.  
These approximations made the reconstruction computationally simpler.  However, eliminating these approximations would likely improve the quality of the reconstruction.  We hope to explore these improvements in future work.

\bibliographystyle{unsrt}
\bibliography{references}

\end{document}